\documentclass[letterpaper, 11pt]{IEEEtran}
\usepackage[margin=1.0in]{geometry}
\IEEEoverridecommandlockouts
\usepackage{soul}
\usepackage{xcolor}
\sethlcolor{yellow}
\usepackage{graphicx}
\usepackage{textcomp}
\usepackage{multirow}

\def\showtrack{0}   %Show deleted changes
\def\showchange{0}  %highlight new changes

\if\showtrack1
    \usepackage{ulem}
    \newcommand\redsout{\bgroup\markoverwith{\textcolor{red}{\rule[0.5ex]{2pt}{0.8pt}}}\ULon}
\else
    \newcommand{\redsout}[1]{}
\fi

\if\showchange1
    
\else
    
\fi
\def\Rtrack{0}   %Show deleted changes
\def\Rchange{0}  %highlight new changes

\if\Rtrack1
    \usepackage{ulem}
    \newcommand\Rredsout{\bgroup\markoverwith{\textcolor{red}{\rule[0.5ex]{2pt}{0.8pt}}}\ULon}
\else
    \newcommand{\Rredsout}[1]{}
\fi

\if\Rchange1
    
\else
    
\fi
\begin{document}

\title{DeFakePro: Decentralized DeepFake Attacks Detection using ENF Authentication}

\author{
\IEEEauthorblockN{Deeraj Nagothu${^a}$, Ronghua Xu${^a}$, Yu Chen${^a}$, Erik Blasch${^b}$, Alexander Aved${^b}$}
\IEEEauthorblockA{\\${^a}$Dept. of Electrical and Computer Engineering,
Binghamton University, Binghamton, USA \\
${^b}$The U.S. Air Force Research Laboratory, Rome, USA\\
Emails: \{dnagoth1, rxu22, ychen\}@binghamton.edu, \{erik.blasch, alexander.aved\}@us.af.mil}
}
\maketitle

\begin{abstract}
Advancements in generative models, like Deepfake allows users to imitate a targeted person and manipulate online interactions. It has been recognized that disinformation may cause disturbance in society and ruin the foundation of trust. This article presents DeFakePro, a decentralized consensus mechanism-based Deepfake detection technique in online video conferencing tools. Leveraging Electrical Network Frequency (ENF), an environmental fingerprint embedded in digital media recording, affords a consensus mechanism design called Proof-of-ENF (PoENF) algorithm. The similarity in ENF signal fluctuations is utilized in the PoENF algorithm to authenticate the media broadcasted in conferencing tools. By utilizing the video conferencing setup with malicious participants to broadcast deep fake video recordings to other participants, the DeFakePro system verifies the authenticity of the incoming media in both audio and video channels. 
\end{abstract}

\section{Introduction}
\label{sec:intro} 

The rise of the fifth generation (5G) communication and the Internet of Video Things (IoVT) technologies enables a broader range of applications with mega-scale data (e.g., all conditions all-time video), while COVID-19 forces more activities like meetings and conferences migrated to the cyberspace. While these network-based applications become essential in the \emph{new normal}, which highly depend on a reliable, secure, real-time audio or/and video streaming (e.g., Zoom), they become a target for attackers \cite{mehta2021fakebuster}. Enhanced with such security features, users tend to rely on the communication channel for confidential conversations and have a higher trust factor on the information received through audio or video mediums. Hence, end-to-end multimedia attacks have a significant impact where the perpetrator is a trusted participant in the conference who can relay misinformation \cite{kagan2020zooming}.

\begin{figure}[t]
    \begin{center}
    \includegraphics[width=0.42\textwidth]{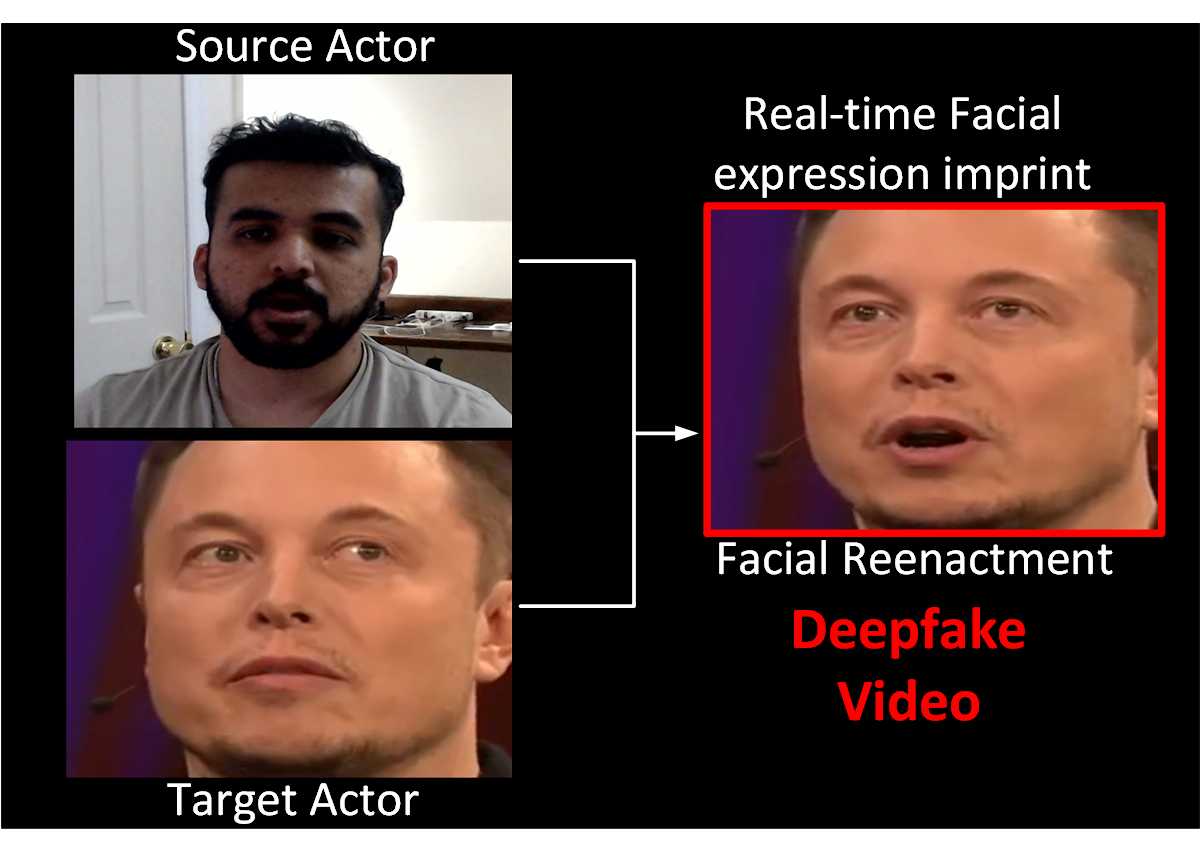}
    \vspace{-0.15in}
    \caption { \label{fig:deepfake_attack} Facial Re-enactment DeepFake attack.} 
    \end{center}
    \vspace{-0.3in}
\end{figure}

Modern generative deep learning models have enabled forging audio and video recordings with another source and creates false media called Deepfakes \cite{goodfellow2014generative}. The Deepfakes are a more potent form of visual layer attacks since it involves manipulating the video and audio channels by imitating a targeted person's face and voice and creating a false recording to relay misinformation through a forged and trusted entity \cite{frank2020leveraging}. Generating such recordings is not difficult with the vast availability of source images and video recordings over the Internet \cite{verdoliva2020media}. Recent advancements in audio software called Descript allow a user to generate text-to-speech content with training data within 10 minutes \cite{DescriptCreatePodcasts}. Deepfaked video, audio or photos in social media are highly disturbing and able to mislead the public, raising further challenges in policy, technology, social, and legal aspects \cite{westerlund2019emergence, durall2020watch}. Figure \ref{fig:deepfake_attack} shows an example of a deepfake attack on a celebrity mimicking the source actor.

Electrical Network Frequency (ENF) is a unique environmental fingerprinting technique for real-time distributed authentication \cite{xu2021decentralized}. ENF is an instantaneous frequency in the power distribution networks, and the fluctuations occur due to the load control variations. ENF is embedded in multimedia recording from different power sources, and the resultant media can be authenticated based on the time stamp and ENF fluctuation patterns recovered from it \cite{grigoras2009applications}. The similarity of the fluctuations along with its robust and random nature makes ENF a reliable source for authenticating digital media recordings. Existing ENF-exploited solutions rely on centralized architectures that can be a performance bottleneck and vulnerable to a single point of failure \cite{nagothu2019detecting}.

Blockchain utilizes a decentralized architecture, which mitigates the problem of a single point of failure and allows for immutable data storage and verification. With the consensus protocol, blockchain executes transactions on a public distributed ledger allow for transparency, immutability, and auditability ensuring data authenticity among untrusted devices. Thus, Blockchain is promising to enable a decentralized authentication mechanism for ENF-based deepfake detection \cite{nikouei2018real}.

Inspired by spatio-temporal sensitive ENF contained in multimedia signals and decentralized consensus algorithms in blockchain, this article proposes DeFakePro, a novel decentralized ENF-consensus-based deepfake detection in audio-video channels for online conferencing scenarios. DeFakePro contributes the following:
 
 \begin{itemize}
     \item The embedded ENF fingerprints of Deepfaked audio and video streams are studied;
     
     \item DeFakePro, a secure deepfake authentication system is introduced along with details of key components and workflows;
     
     \item A partially decentralized PoENF consensus algorithm is designed to ensure the efficiency and security in distributed authentication of streaming media; 
     
     \item A proof-of-concept prototype is tested with deepfaked audio and video authentication in an online video conferencing setup and verifies the feasibility of the DeFakePro system; and 
     
     \item An experimental evaluation of the proposed system with the current state-of-the-art technique shows that DeFakePro has similar accuracy performance, but comparatively faster making it suitable for online applications.
     
 \end{itemize}

\section{Deepfake Attacks on Online Conferencing Tools}

With advanced computation power and development in deep learning (DL) models, generative adversarial networks (GAN) can imprint the source facial landmarks or impressions on a targeted person to recreate similar content with a fake personality commonly referred to as Deepfake \cite{goodfellow2014generative}. Both audio and video recordings can be manipulated with enough training data available from the source. Current Deepfake detection relies on DL models trained to detect visual artifacts introduced in the deepfake videos. However, with more training data and models, such artifacts can be removed, and deepfake videos' precision gets better \cite{verdoliva2020media}.

For online digital media in the context of conference tools, both audio and video are equally targeted to create the mirage of fake digital presence. Authentication of both audio and video recording for forgery detection is eminent for information integrity \cite{zhou2021joint} in all available channels. A detection technique void of training data and large scale computational infrastructure, which depends on underlying fingerprints or multimedia artifacts to locate deepfake forgeries, enables reliable digital media authentication. DeepFakes can perform better with more training data and create visually perfect manipulation of the target, however it results in high frequency artifacts and shows poor performance in reconstructing spectral consistencies \cite{durall2020watch,frank2020leveraging}. 

Application with simple video manipulations like Face-swap or Face Shifting software and audio manipulation like generating text-to-audio speech on the go using a modified voice have become abundantly available for common users \cite{DescriptCreatePodcasts, perov2020deepfacelab}. For online conferencing tools, the participating perpetrator has complete control over the audio and video broadcasted to other users. With such manipulation tools easily accessible, the perpetrator can imitate a targeted person and spread misinformation. Such attacks raise concerns over the virtual communication platforms, and along with the existing network-level security, online conferencing software also requires an authentication scheme for the information broadcasted. 

A study in video deepfake detection using the spatial frequency inconsistencies caused by the up-sampling mechanism of most deepfake generator models confirms the frequency-level modifications \cite{durall2020watch}. The resulting frequency fingerprints are utilized to train neural networks to detect the GAN-based modifications \cite{frank2020leveraging,marra2019gans}. However, the spatial frequency inconsistency still remains a trainable parameter and the resulting fingerprints could be minimized \cite{durall2020watch}. Leveraging the incapability of DeepFake models to preserve spectral consistencies and the random nature of ENF consisting of spatio-temporal fingerprint information, we introduce DeFakePro channels as a distributed authentication system for online media broadcasts. The proposed DeFakePro identifies fingerprints sensitive to both spatial and temporal frequency manipulations.

% ------------------------------------------------------------------------------
\section{DeFakePro System}

\begin{figure}[t]
    \begin{center}
    \includegraphics[width=0.43\textwidth]{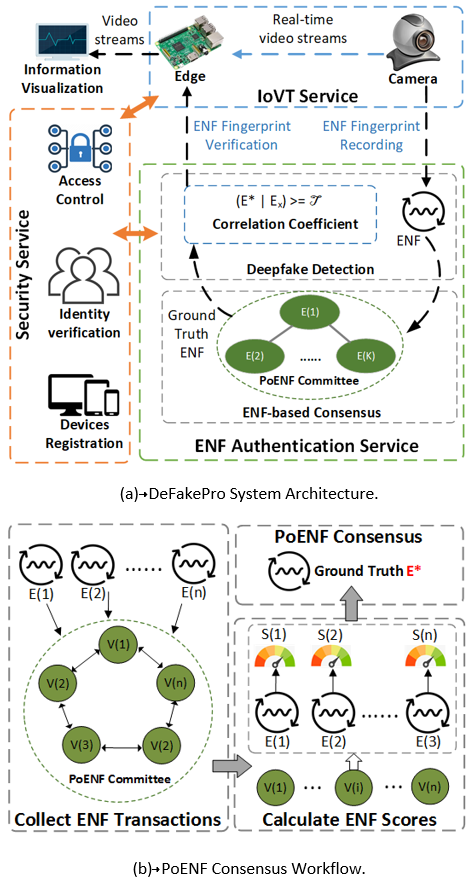}
    \vspace{-0.15in}
    \caption { \label{fig:architecture} DeFakePro System.} 
    \end{center}
    \vspace{-0.3in}
\end{figure}

The DeFakePro system comprises of a decentralized authentication system, where each participating node estimates ENF and broadcasts it for proof of authenticity. Figure \ref{fig:architecture}(a) represents the system workflow. Each modules are discussed as follows. 

\subsection{ENF Estimation in Multimedia Recordings}

The presence of an ENF signal in a multimedia recording depends on the location of the recording, where the estimation of the signal varies based on the type of multimedia source. DeFakePro tackles both audio and video streams for dual authentication and adds robustness to the system. For the online conferencing setup, we assume the recordings are made in an indoor environment.

\subsubsection{ENF in Audio Recordings}
The ENF is embedded in an audio recording through electromagnetic induction for devices powered by the electrical grid \cite{grigoras2009applications} or background hum for devices running on a battery-powered source \cite{chai2013source}. Audio recordings are typically captured with a high sampling frequency (44.1 kHz), but for ENF estimation, a low frequency like 1kHz - 8kHz is sufficient.  

\subsubsection{ENF in Video Recordings} 
The source of ENF in video recordings is through the illumination frequency from light sources powered by the electrical grid. The illumination frequency, i.e., 120Hz when the nominal frequency is 60Hz, is captured by the recording devices based on the imaging sensor used \cite{garg2013seeing}. The two most commonly used imaging sensors are complementary metal-oxide semiconductor (CMOS) and charge-coupled device (CCD) sensors. Each sensor has its unique shutter mechanism associated with image capturing, and the total samples captured depend on the number of frames per second (FPS). CCD sensors utilize a global shutter mechanism where the whole sensor is exposed to light for each frame, resulting in lower samples. In the case of CMOS sensors, the frames are captured using a rolling shutter mechanism, where each row of the imaging sensor is sequentially exposed to light, and the number of ENF samples captured is increased by the frame height \cite{vatansever2019analysis}. Among CCD and CMOS-based sensors, CMOS is most commonly used for general purposes due to its cost efficiency and broad applicability.

\subsubsection{ENF Estimation} 
The ENF is estimated in the following steps:

\begin{itemize}
\item[i)] \emph{Power Spectrum Matrix} is computed using the spectrogram technique from the collected samples in audio and video recordings;
\item[ii)] Based on the \emph{Nominal frequency}, the weights are estimated from the harmonic frequencies in the power spectrum matrix; and
\item[iii)] The computed weights are used to \emph{Combine Spectrum Slices} resulting in a robust ENF estimation.
\end{itemize}

The detailed discussion of ENF estimation techniques for different multimedia recordings are described in our previous work \cite{nagothu2021authenticating, nagothu2019detecting}. 

\subsection{Security Service}
DeFakePro leverages security services to provide basic cryptography security primitives for permissioned Internet of Video Things (IoVT), as shown in the left of Fig. \ref{fig:architecture}. All devices and users must complete registration to join the network, and DeFakePro assumes that a system administrator is a trust oracle to manage profiles of all registered entities. DeFakePro relies on container technology to implement security services that support resource isolation, data flexibility, and maintenance simplicity in a distributed network environment. Each service unit exposes a set of RESTful web-service APIs for devices/users. Identity verification services rely on a virtual trust zone method to authenticate identities. Access control services use a capability-based access model to support decentralized access authorization and verification \cite{xu2021decentralized}.

\subsection{Proof-of-ENF (PoENF) Consensus}

To maintain ground truth ENF benchmarks used for deepfake detection, the DeFakePro solution designed a byzantine resistant PoENF consensus algorithm that is executed by a PoENF committee. Such a committee can be either pre-configured by a system administrator or random elected given a certain period of time. Figure \ref{fig:architecture}(b) illustrates the PoENF consensus workflow consisting of three main procedures.

\subsubsection{Collect ENF Transactions}
At the beginning of current consensus round, a validator $V(i)$ can broadcast a ENF transaction saving its ENF proof $E(i)$ among PoENF committee members. Then, other validators can verify a received ENF transaction given conditions that a) it should be sent by validators in committee; and b) it should be neither outdated or existed in local transactions pool. Finally, all valid transactions are locally buffered.  

\subsubsection{Calculate ENF Scores}
Given a local ENF transaction pool, a validator $V(i)$ can extract ENF proofs from other committee members and build a global view of collected ENF proofs. To prevent against byzantine validators who send arbitrary or poisoned ENF proofs, DeFakePro adopts a byzantine resilient aggregation rule in the ENF score calculation. Finally, each validator has a global view of ENF scores, as shown by Figure \ref{fig:architecture}(b).

\subsubsection{PoENF Consensus}
In the PoENF consensus stage, every validator can sort ENF scores and choose the minimum one as ground truth $E^*$. As all honest validators have the identical global view of ENF scores, they can generate the same $E^*$. The PoENF requires that a validator always uses $E^*$ as the ground truth ENF. Therefore, PoENF consensus can make an agreement on $E^*$ given an assumption that an adversary can only compromise at most $f$ committee members.

Interested readers can refer to \cite{xu2021econledger} for details about PoENF consensus protocol. 

\subsection{Deepfake detection using PoENF Consensus}
Once the PoENF consensus agrees on the ground truth ENF $E^*$ for the round, each node compares its local ENF with the ground truth ENF using the correlation coefficient. The measure of similarity ranges from $[-1,1]$, where $1$ represents highest similarity. Based on the experiments, we adopted a threshold of $0.8$ to compare the ENF signals. For localization of the forgery, a sliding window protocol is used to compare the ENF signal.

% ------------------------------------------------------------------------------
\section{Prototype and Evaluation}

\subsection{Experimental Setup}

A proof-of-concept prototype of DeFakePro is implemented in python. To emulate the participants in an online conferencing tool, we adopted Raspberry Pi-4 (RPi) as the nodes to cap the computation power requirements. For performance of PoENF consensus, we compared the time latency on RPi and a Dell Optiplex-7010 desktop. The collected raw footage is processed in the devices and using the ENF estimation techniques, the ENF signal is broadcasted to the PoENF committee. For Audio deepfakes, the \emph{Descript} platform \cite{DescriptCreatePodcasts} is used, where the software can generate a text-to-speech synthesis in real-time for any pre-trained vocals. For Video Deepfakes, a live deepfake generator named \emph{DeepFaceLive} is used \cite{perov2020deepfacelab}. The video deepfake generator uses the live webcam feed, and synthesizes the targeted users face on the source image. 

\subsection{Performance of PoENF consensus mechanism}
Table \ref{tab:latency} presents the cumulative time taken for a round of PoENF consensus, including ENF proof broadcast, verification, and PoENF algorithm execution. 
The time complexity of the PoENF consensus is $\mathcal{O}(K^2d)$, where $K$ is the committee size, $d$ is the ENF sample size, and the latency increases with the number of validators. A general conference scenario typically includes less than 50 participants and incurs delays up to $0.5$ sec and $0.2$ sec on a desktop. 

\begin{table}[t]
% \vspace{-0.10in}
\caption{PoENF consensus latency (Second) with different number of validators. Comparative evaluations on platform benchmarks} 
\vspace{-0.18in}
\label{tab:latency}
\begin{center}       
\begin{tabular}{|p{1.4cm}|c|c|c|c|c|c|c|c|c|c|} %% this creates two columns
%% |l|l| to left justify each column entry
%% |c|c| to center each column entry
%% use of \rule[]{}{} below opens up each row
\hline
\rule[-1ex]{0pt}{3.5ex} \textbf{No. of Validators} & 10 & 20 & 50 & 100 & 200 & 500 \\
\hline
\rule[-1ex]{0pt}{3.5ex} \textbf{RPi-4} & 0.02 & 0.08 & 0.48 & 1.93 & 7.71 & 48.5 \\
\hline
\rule[-1ex]{0pt}{3.5ex} \textbf{Desktop} & 0.01 & 0.02 & 0.15 & 0.59 & 2.4 & 15.4 \\
\hline
\end{tabular}
\end{center}
\vspace{-0.20in}
\end{table}

\subsection{Detecting Audio and Video Deepfakes}

\begin{figure*}[t]
    \begin{center}
    \includegraphics[width=1\textwidth]{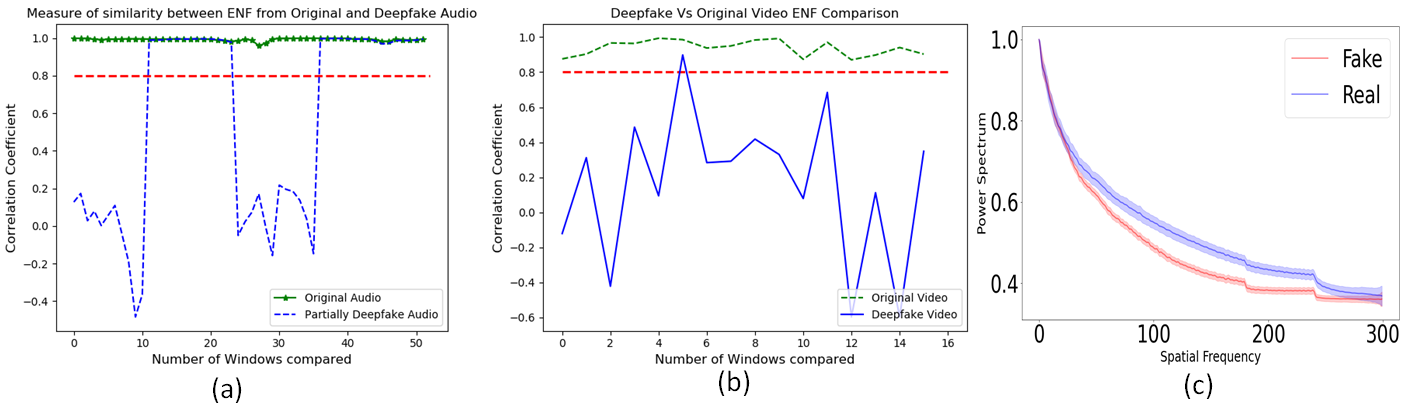}
    \vspace{-0.15in}
    \caption { \label{fig:deepfake_experiments} (a) Original and Deepfaked audio recording correlated with ground truth ENF (b) Real-time face swap Deepfake generated using DeepFaceLab software and compared with ground truth ENF (c) Spectral inconsistencies generated by Deepfake model, generated using Frequency transform and Azimuthal averaging.} 
    \end{center}
    \vspace{-0.3in}
\end{figure*}

The audio and the video deepfakes are generated independently to analyze the effects of ENF on each recording. For audio recording broadcasted, the text-to-speech modification is made in multiple locations throughout the recording \cite{DescriptCreatePodcasts}. After comparing the ENF from the ground truth ENF and the audio deepfake ENF with multiple forgery locations, a low correlation coefficient indicates a fake audio from which the modified section can be localized. Figure \ref{fig:deepfake_experiments}(a) represents the mismatch in the ENF, where the correlation coefficient is below the threshold.

\begin{figure}[t]
    \begin{center}
    \includegraphics[width=0.48\textwidth]{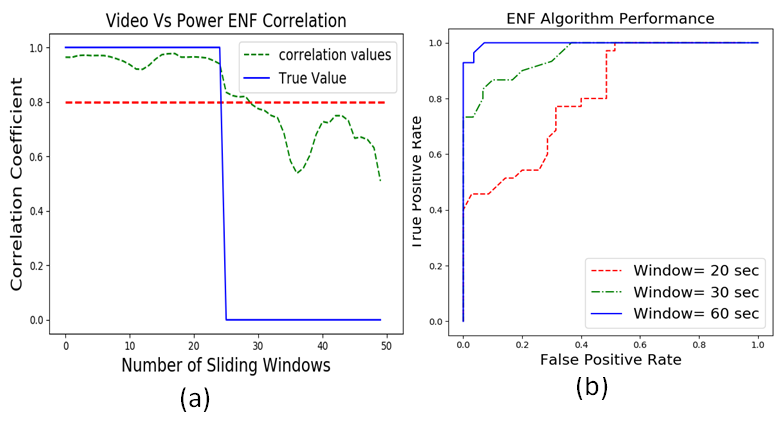}
    \vspace{-0.15in}
    \caption { \label{fig:roc_compare} (a) Correlation coefficient values for Deepfake localization (b) ROC curve for optimal ENF window size for lower false positives and threshold selection.} 
    \end{center}
    \vspace{-0.3in}
\end{figure}

For video recordings, the deepfakes are generated in real-time using a pre-trained model face set in the DeepFaceLab tool \cite{perov2020deepfacelab}. For our experiment, multiple deepfake videos were generated using the DeepFaceLab tool. For the generated deepfake videos, a subset of frames from each video is analyzed for spatial frequency inconsistencies generated due to the deepfake model up-sampling mechanism \cite{durall2020watch}. Figure \ref{fig:deepfake_experiments}(c) represents the changes in spatial frequencies caused by most deepfake models, generated by analyzing the azimuthally averaged frequency spectrum of deepfake frames \cite{durall2020watch}. Along with the facial manipulations in the frame center, the spatial frequency inconsistencies represent that the deepfake model adds additional perturbations in the static background of the frame.

The video frames are buffered in online conferencing tools to collect enough samples for a reliable ENF estimation. With more samples, ENF estimation is more accurate. A sliding window approach is used for an online authentication system to buffer incoming frames and estimate ENF. We tested various window sizes and a fixed shift size of $5$ seconds for our experiment since shift size has a low effect on ENF estimation. Figure \ref{fig:roc_compare}(a) shows a clear separation between original frames and deepfake frames, while Figure \ref{fig:roc_compare}(b) represents the accuracy of detecting deepfake videos as window sizes vary.

Using the appropriate window and shift sizes, ENF-based video authentication is presented in Figure \ref{fig:deepfake_experiments}(b). Given the ground truth ENF, the measure of similarity of the incoming deepfake video ENF estimates is lower than the original video streams. In deepfake recordings, even though the facial landmark regions are forged, the pixel intensities through the frame are modified due to added perturbations, as seen in Figure \ref{fig:deepfake_experiments}(c).

\subsection{Performance Evaluation}

A comparison study is performed to analyze the effectiveness of ENF-based authentication compared to the spatial frequency-based GAN fingerprint. To the best of our knowledge, the presented approach is the only technique focused on online deepfake detection using a distributed backbone system. We collected ten 5-minutes Deepfake videos, collected with ground truth ENF, and evaluated the performance using ENF-technique and spatial inconsistencies based detection \textit{UpConv} in \cite{durall2020watch}.

\begin{table}[t]
% \vspace{-0.10in}
\caption{DeepFake Detection Performance Comparison} 
\vspace{-0.18in}
\label{tab:performance_comparison}
\begin{center}       
\begin{tabular}{|c|p{0.5 cm}|p{0.5 cm}|p{0.5 cm}|p{0.5 cm}|p{0.5 cm}|p{0.5 cm}|} %% this creates two columns
%% |l|l| to left justify each column entry
%% |c|c| to center each column entry
%% use of \rule[]{}{} below opens up each row
\hline
\rule[-1ex]{0pt}{3.5ex} \multirow{2}{*}{\textbf{Techniques}} & \multicolumn{2}{c|}{1080p} & \multicolumn{2}{c|}{720p} & \multicolumn{2}{c|}{480p} \\\cline{2-7}
\rule[-1ex]{0pt}{3.5ex}  & AUC & FPS & AUC & FPS & AUC & FPS\\
\hline
\rule[-1ex]{0pt}{3.5ex} \textbf{DeFakePro} & 0.95 & 19 & 0.96 & 25 & 0.95 & 33 \\
\hline
\rule[-1ex]{0pt}{3.5ex} \textbf{UpConv} & 0.94 & 10 & 0.97 & 14 & 0.98 & 20 \\
\hline
\end{tabular}
\end{center}
\vspace{-0.20in}
\end{table}

Table \ref{tab:performance_comparison} represents the performance of each Deepfake detection technique in multiple resolutions. For online conference scenarios, faster and reliable techniques are more viable due to its time sensitive nature. The Area under the Curve (AUC) for both techniques is similar for all formats, however the number of Frames per second (FPS) for the proposed DeFakePro system is higher since there is minimal frame processing required and no feature training. DeFakePro is applicable to any input streams as long as it carries background ENF signature, and the nominal frequency is known. The presented approach is effective against any kind of frame modification since the ENF fingerprint carries unique fluctuations and allows for a distributed authentication system enabling Deepfake detection and byzantine nodes.

\section{Conclusions}

This article presents DeFakePro - a decentralized deepfake attack detection system leveraging embedded ENF signals in online video conferencing tools. The proposed DeFakePro adds resilience to byzantine nodes and verifies media integrity with minimal computational resources using the integrated PoENF consensus mechanism. The consensus mechanism establishes the ground truth ENF in each round, and each participating node can verify the media authenticity using a correlation coefficient. Furthermore, the consensus mechanism is evaluated for time latency based on the number of participants in each round. However, the application of ENF-based authentication is limited to zones with passive ENF presence, like indoor environments.

The experimental results show that the DeFakePro system can detect and localize the deepfake audio and video attacks using the estimated ENF signal. The DeFakePro system is evaluated against the current Deepfake detection techniques, and the proposed system achieves similar performance and had faster processing rate which is pre-requisite for an online detection system.

\section*{Acknowledgement}

This work is supported by the U.S. National Science Foundation (NSF) via grant CNS-2039342 and the U.S. Air Force Office of Scientific Research (AFOSR) Dynamic Data and Information Processing Program (DDIP) via grant FA9550-21-1-0229. The views and conclusions contained herein are those of the authors and should not be interpreted as necessarily represenSting the official policies or endorsements, either expressed or implied, of the U. S. Air Force. 

% References
\bibliographystyle{IEEEtran} % makes bibtex use IEEEtran.bst
\bibliography{References} % bibliography data in References.bib

\vskip -2\baselineskip plus -1fil

%% Include maximum *60 word* bio of each author with email ID *after* the references list.
\begin{IEEEbiographynophoto}{Deeraj Nagothu}
is a PhD candidate of electrical and computer engineering at the Binghamton University - SUNY. He received his MS degree on electrical and computer engineering from Binghamton University in 2016. His current research interests are multimedia forensics in internet of video things (IoVT) and computer network security.
\end{IEEEbiographynophoto}
\vskip -2.8\baselineskip plus -1fil
\begin{IEEEbiographynophoto}{Ronghua Xu}
is a PhD candidate of electrical and computer engineering at Binghamton University. He earned a BS in mechanical engineering from Nanjing University of Science and Technology, China, in 2007, and a MS on mechanical and electrical engineering from Nanjing University of Aeronautics and Astronautics in 2010. His research interests are blockchain based security solutions to internet of things (IoT).
\end{IEEEbiographynophoto}
\vskip -2.8\baselineskip plus -1fil
\begin{IEEEbiographynophoto}{Yu Chen}
is an Associate Professor of electrical and computer engineering at Binghamton University. He received a PhD in Electrical Engineering from the University of Southern California (USC) in 2006. His research interest lies in edge-fog-cloud computing, IoTs, and smart cities. He published over 200 papers in journals and conference proceedings. He is a senior member of IEEE and SPIE, and a member of ACM.
\end{IEEEbiographynophoto}
\vskip -2.8\baselineskip plus -1fil
\begin{IEEEbiographynophoto}{Erik Blasch} 
is with the Air Force Research Laboratory (AFRL). He received his BS degree from MIT and his PhD from Wright State University (1999) in addition to seven master’s degrees. He has been with the AFRL since 1996, compiling over 900 papers, 42 patents, and 8 books. He is a Fellow of IEEE, AIAA, SPIE, and MSS. His research areas include target tracking, image fusion, information fusion performance evaluation, and human-machine integration.
\end{IEEEbiographynophoto}
\vskip -2.8\baselineskip plus -1fil
\begin{IEEEbiographynophoto}{Alex J. Aved} 
received the BA degree in Computer Science and Mathematics in 1999 from Anderson University in Anderson, Indiana, an MS in Computer Science from Ball State University and PhD in Computer Science in 2013 from the University of Central Florida. He is currently a technical advisor at the Air Force Research Laboratory Information Directorate in Rome, NY. Alex’s research interests include multimedia databases, stream processing and dynamically executing models with feedback loops incorporating measurement and error data to improve the accuracy of the model.
\end{IEEEbiographynophoto}

\end{document}